\documentclass[sigconf,10pt]{acmart}
\AtBeginDocument{%
  \providecommand\BibTeX{{%
    \normalfont B\kern-0.5em{\scshape i\kern-0.25em b}\kern-0.8em\TeX}}}

\copyrightyear{2022} 
\acmYear{2022} 
\setcopyright{acmlicensed}\acmConference[ICS '22]{2022 International Conference on Supercomputing}{June 28--30, 2022}{Virtual Event, USA}
\acmBooktitle{2022 International Conference on Supercomputing (ICS '22), June 28--30, 2022, Virtual Event, USA}
\acmPrice{15.00}
\acmDOI{10.1145/3524059.3532389}
\acmISBN{978-1-4503-9281-5/22/06}

\usepackage{graphicx}
\usepackage{listings}
\usepackage{subcaption}
\usepackage{tabularx}
\usepackage{wrapfig}
\usepackage{booktabs}
\usepackage{multirow}
\usepackage{tikz}
\usepackage{pgfplots}
\usepackage{dblfloatfix}
\usetikzlibrary{arrows,shapes,positioning,shadows,trees}

\graphicspath{ {./img/} }

\definecolor{shadecolor}{RGB}{250, 250, 250}
\definecolor{darkgreen}{RGB}{0,127,0}
\definecolor{darkred}{RGB}{127,0,0}
\begin{document}

\title{Lifting C Semantics for Dataflow Optimization}

\author{Alexandru Calotoiu}
\email{acalotoiu@inf.ethz.ch}
\affiliation{\institution{ETH Zurich, Switzerland}}
  
\author{Tal Ben-Nun}
\email{talbn@inf.ethz.ch}
\affiliation{\institution{ETH Zurich, Switzerland}}
\author{Grzegorz Kwasniewski}
\email{gkwasnie@inf.ethz.ch}
\affiliation{\institution{ETH Zurich, Switzerland}}
\author{Johannes de Fine Licht}
\email{definelj@inf.ethz.ch}
\affiliation{\institution{ETH Zurich, Switzerland}}
\author{Timo Schneider}
\email{timo.schneider@inf.ethz.ch}
\affiliation{\institution{ETH Zurich, Switzerland}}
\author{Philipp Schaad}
\email{philipp.schaad@inf.ethz.ch}
\affiliation{\institution{ETH Zurich, Switzerland}}
\author{Torsten Hoefler}
\email{torsten.hoefler@inf.ethz.ch}
\affiliation{\institution{ETH Zurich, Switzerland}}

\renewcommand{\shortauthors}{Calotoiu, et al.}

\begin{abstract}
C is the \emph{lingua franca} of programming and almost any device can be programmed using C. However, programming modern heterogeneous architectures such as multi-core CPUs and GPUs requires explicitly expressing parallelism as well as device-specific properties such as memory hierarchies.  The resulting code is often hard to understand, debug, and modify for different architectures.  We propose to lift C programs to a parametric dataflow representation that lends itself to static data-centric analysis and enables automatic high-performance code generation.
We separate writing code from optimizing for different hardware: simple, portable C source code is used to generate efficient specialized versions with a click of a button. 
Our approach can identify parallelism when no other compiler can, and outperforms a bespoke parallelized version of a scientific proxy application by up to 21\%.
\end{abstract}

\ccsdesc[500]{Software and its engineering~Source code generation}
\ccsdesc[500]{Computing methodologies~Parallel algorithms}
\keywords{parallelism, dataflow analysis, automatic parallelization}


\maketitle

\section{Introduction}

Many performance critical applications are written in C, as its machine model is usually closest to hardware and allows for bare-metal tuning to achieve highest performance.
According to the TIOBE index~\cite{tiobenov20} in 2020, C was the most popular language in Internet searches. High-performance computing centers state that 25\% of their users primarily use C~\cite{decscs}.
Since Kernighan's and Ritchie's original inception of the C language, systems have changed dramatically. 
%
%
Most architectures need specialized instructions, compiler directives, or libraries to be used efficiently. This usually leads to C programs where more lines of code are implementing optimizations tailored to the architecture than solving the actual problem.

Targeted optimization is tightly coupled to hardware architectures. A code written for GPUs using CUDA, a code written to exploit shared memory using OpenMP, and a code written for large supercomputers using the message passing interface (MPI) can be nominally written in C, but will vary widely even if they solve the same problem. 
The only aspect they are likely to have in common is the  sequential algorithm each variant is based on.
We argue that specializing the programs to an architecture treats the symptoms, but cannot eliminate the root cause: precisely because C was \emph{not} designed for \emph{performance portability}, optimizing C programs is both challenging and time consuming. 
\begin{figure*}[t]
    \centering
    \includegraphics[width=1.0 \linewidth,page=2,clip,trim={0cm 16.39cm 13.3cm 0cm}]{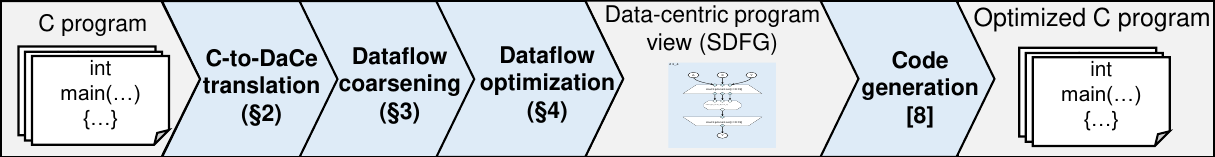}
    \caption{Optimizing C programs by lifting dataflow.}
    \label{fig:analysis}
\end{figure*}

A powerful alternative to specialization is using tools provided by modern compilers such as polyhedral analysis~\cite{polly,pluto} to optimize and parallelize sequential C code, with results rivaling and even surpassing hand-tuned versions of the code. However, these are limited to static control parts (SCOPs) within functions~\cite{polly}. SCOPs impose constraints on what type of source code can be analyzed: indirect array accesses such as $x[column\_index[j]]$  are typically not permitted. The limitation is apparent in the following example (sparse matrix vector multiplication), as no optimization is possible due to the data-dependent indirect array accesses.
\begin{lstlisting}[language=C,aboveskip=0pt,belowskip=0pt]
for (i = 0; i < N; i++)
    for (j = row_ptr[i]; j < row_ptr[i + 1]; j++)
        y[i] += A[col_idx[j]] * x[j];
\end{lstlisting}
In search of a more general solution, we observe that data movement is \emph{the} most expensive part of most program executions when considering both energy and time~\cite{ivanov20}.
Data-centric programming and leveraging dataflow graphs is already widely performed in compiler analysis~\cite{llvm,hpvm,bamboo}, and recently emerging in graph analytics~\cite{gunrock}, high performance computing~\cite{legion,maps}, and machine learning \cite{tensorflow}.

Data-centric models are both productive and portable, as parallelism is inherently expressed as data-independent sections, regardless of the target hardware. 

Our goal is to generate optimized, parallel code for different platforms by minimizing data movement.
 To achieve it, we \emph{extract the data movement semantics from most C programs into a parametric dataflow representation}, where data movement can be better analyzed and transformed.  

While one cannot statically analyze the dataflow of all C programs, as can be shown by the Halting problem or Rice's theorem, we observe that high performance C codes, a subset of C programs without undefined behavior, recursion or function pointers, can be lifted.
%

%

\begin{figure*}[!t]
    \centering
    \includegraphics[width=1.0 \linewidth,page=1,clip,trim={0cm 6.1cm 1cm 0cm}]{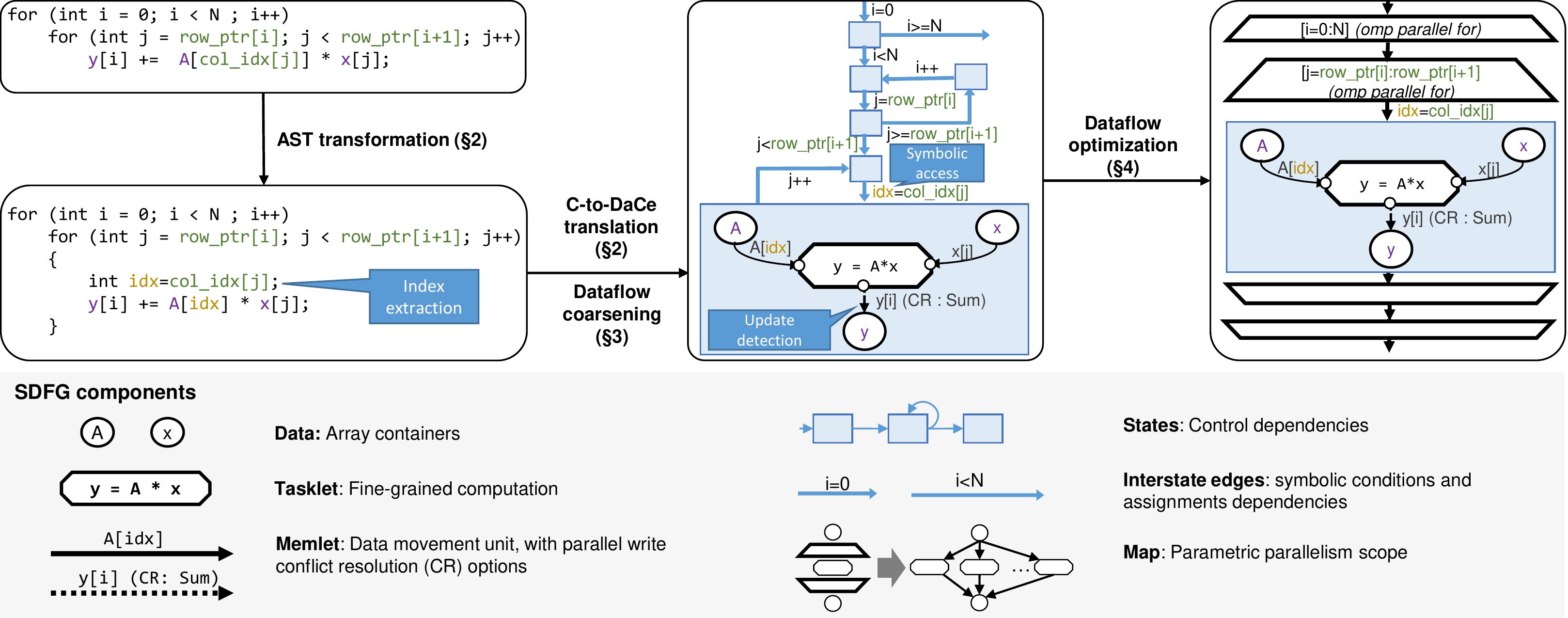}
    \caption{From C to Data-Centric Programming.}
    \label{fig:sdfg}
\end{figure*}
We keep track of memory accesses using symbolic analysis of access patterns and leverage the dataflow across the entirety of a program. To showcase the opportunities provided by the data-centric approach we show how we can automatically expose data parallelism by identifying and optimizing updates to shared memory locations. We evaluate the effectiveness of our parallelization by providing an automatic method of deriving work/depth models for code we have parallelized.
There is no need to annotate the code to recover parametrically-parallel sections, as we derive the required information directly from dataflow. The overall process is fully automatic and is summarized in Figure~\ref{fig:analysis}. Figure~\ref{fig:sdfg} shows a more detailed view of how the sparse matrix vector multiplication code is translated, transformed and optimized and will be discussed in detail in Sections~\ref{sec:frontend},~\ref{sec:extraction}, and~\ref{sec:optimizing}.

As we shall show, from the raw C codes, we are able to not only generate codes that perform equivalently or better than specialized tools such as polyhedral compilers; but also operate on LULESH~\cite{lulesh}, a scientific computing application, finding parallelization opportunities that no state-of-the-art tool detects, and even \textit{outperform the tuned parallel version provided by the application authors}.

\paragraph{Contributions} 
\begin {itemize}
\item We statically lift the semantics of dataflow from C into a data-centric intermediate representation.\footnote{Prototype available at: \url{https://github.com/spcl/c2dace}.}
\item We use symbolic analysis of data access patterns across entire programs to expose optimizations and parallelism in unmodified C programs.
\item We statically detect the update of a memory location as a distinct data access pattern to expose additional parallelism opportunities.
\item We introduce an automatic, static work-depth analysis to objectively measure the degree to which we have exposed parallelism in sequential C code.
\item On the LULESH~\cite{lulesh} high-performance scientific application, we automatically generate a parallel version that outperforms all other compilers and autoparallelizing tools and even surpasses the developers' own OpenMP parallelization by up to 21\%.
\end {itemize}

\section{From C to Data-Centric Programming}\label{sec:frontend}

The data-centric programming paradigm revolves around memory, its movement, and its manipulation through computations. 
Rather than prioritizing control-flow constructs (e.g., sequential statements, loops), the core component of data-centric models is dataflow. Execution order is thus first a byproduct of data dependencies, and secondly a result of explicit control-flow.
There are three governing principles to the paradigm: separation of data containers from computation, explicit data movement expressed as a first-class component, and providing control dependencies for cases where dataflow is not implied (e.g., data-dependent branches).


This is a crucial difference to control-centric C programs, where dataflow is implicit. In order to perform this paradigm shift we must execute a workflow to lift dataflow from C programs. Throughout this workflow, we must maintain semantic equivalence in every step of the translation. We separate the workflow: first, we perform AST transformations to simplify the translation to the dataflow representation. Then, we parse the C code into a fine-grained dataflow representation. Then we repeatedly coarsen that dataflow, after which we can perform optimizing transformation passes. Finally, we can generate optimized C source code for different architectures.

In this work, we focus on the Stateful Dataflow Multigraph (SDFG) IR~\cite{dace} as the data-centric representation. 
An SDFG is a directed graph, representing a state machine, where each node (\textbf{state}) is in itself a parametric directed acyclic multigraph.
In the outer graph, edges contain state transition conditions and assignments. Each state is in turn an acyclic dataflow multigraph, with edges representing data movement and nodes representing data containers, computations, and parametric parallelism scopes. The components are summarized in Figure~\ref{fig:sdfg} and full operational semantics can be found in Ben-Nun et al.~\cite{dace}.

Using the DaCe framework, SDFGs were shown to accelerate a wide range of application classes in dense/sparse linear algebra and graph algorithms~\cite{dace}, deep learning Transformer architectures~\cite{ivanov20}, numerical weather prediction on FPGAs~\cite{stencilflow}, and extreme-scale quantum transport simulations on the world's largest supercomputer~\cite{ziogas19}.


In this work, we introduce a workflow to translate C programs to the SDFG IR. To start, we provide a high level overview mapping major C syntax elements~\cite{csem} to equivalent SDFG elements in Table~\ref{table:c2dace}, and introduce both relevant SDFG components in more detail as well as discuss the more challenging aspects of C to SDFG translation below.

\begin{table}
\footnotesize
\begin{tabular}{p{1.3in} p{1.6in}}
\toprule
 
\bf C Language &	\bf SDFG Equivalent \\
\midrule
\multicolumn{2}{l}{\bf Declarations and Types (\S~\ref{sec:declare})}\\
\midrule
Primitive data type & Scalar data container\\
Array	&Array data container\\
Pointer& Access node to existing data container, or new data container if pointing to newly allocated memory. 
\\
\midrule
\multicolumn{2}{l}{\bf Expressions and Assignments (\S~\ref{sec:assign})}\\
\midrule
Operators (Unary, Binary,...) & Tasklet with incoming and outgoing memlets for read/written operands  \\
Array expression & Memlet\\
\midrule
\multicolumn{2}{l}{\bf Statements (\S~\ref{sec:statement})}\\
\midrule
Compound  (blocks) &	State\\
Branching (\texttt{if,...})&	Branch conditions on state transition edges\\
Iteration (\texttt{for, while, ...)}&	State for compound statement, with states and transitions for loop logic \\
Function flow (\texttt{break, continue, return})	& State transitions\\
\texttt{goto}	& State transition\\ 
\midrule
\multicolumn{2}{l}{\bf Functions   (\S~\ref{sec:func})}\\
\midrule
Function calls (with source)&	Nested SDFG for content, memlets reduce shape of inputs and outputs\\
External/Library calls &	Tasklet with library state 
\\
Recursion & Unsupported\\
Function pointers&	No equivalent, unsupported \\	
\midrule
\multicolumn{2}{l}{\bf Parallelism   (\S~\ref{sec:parallelism})}\\
\midrule
---&	Parametric map scope\\\addlinespace
\bottomrule
\end{tabular}
\caption{Mapping of major C syntax~\cite{csem} elements to SDFG representation.}\vspace{-1em}
\label{table:c2dace}
\end{table}

\subsection{Declarations and Types}
\label{sec:declare}
We need to capture all instances where data is defined, read, and written. The first step is to capture all instances where data is defined, whether statically or at runtime.
The equivalent to declarations in C is the creation of data containers in SDFGs. 

\textbf{Data containers} are accessed using \textit{access} nodes in SDFGs, and represent \textit{arrays}, both one- and multi-dimensional.
Scalars are thus specialized data containers, with just one instance of a primitive data type. 

Some examples of data containers are shown below:
\begin{center}
\includegraphics[width=\linewidth,page=1]{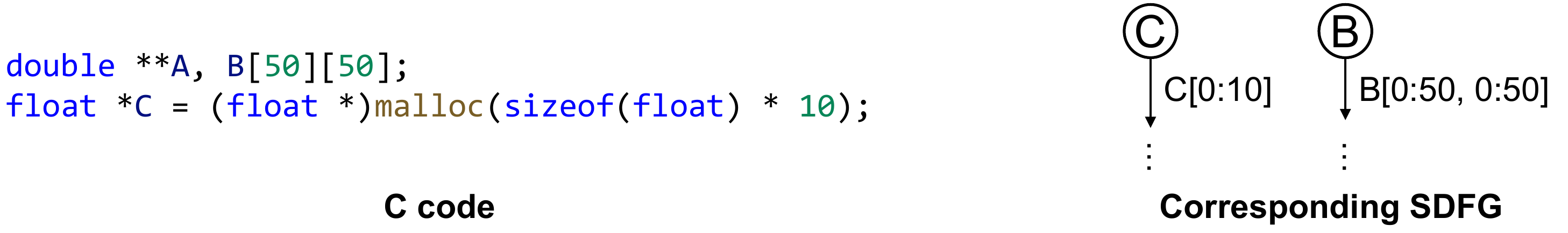}    
\end{center}\vspace{-0.5em}
Here, \texttt{C} will be registered as a one-dimensional single precision floating point array of 10 elements in the SDFG, and \texttt{B} as a two-dimensional double precision floating point array of 50 elements times 50 elements. 
We ensure no aliasing is possible in our representation by not creating a separate data container for pointers such as \texttt{A}. Containers will be created only if \texttt{A} is assigned to newly allocated memory. If \texttt{A} is assigned to an existing data container, \texttt{A} will simply be replaced with an access to that container.

SDFGs rely on symbolic math to perform useful analyses and transformations. A \textbf{symbol} is defined as a scalar that will not be modified within any state. 
Symbols can only be set between states, in an inter-state edge. We can thus use symbolic expressions in memory offsets and integer sets, and differentiate them from runtime-computed scalars.

To analyze dynamically allocated memory such as 
\texttt{malloc} and variable-length arrays, we automatically create symbols out of integer scalar values, as we detail in Section~\ref{sec:scal2sym}.


\subsection{Expressions and Assignments}
\label{sec:assign}

Assignments are some of the most common constructs encountered in C. An assignment contains both data (read and written) and computation (as part of expressions), and we discuss their SDFG equivalents below.

\textbf{Computation} in SDFG is represented by octagonal nodes called \textbf{tasklets}. A tasklet contains C code and may only access memory provided by incoming and outgoing edges. It may not have \textit{side effects} with other computations in the graph, i.e., the operations on one tasklet must not affect computation in any other tasklet, including other instances or invocations of itself. 

Between data and computation, \textbf{data movement} is explicitly represented by edge attributes
called \textit{memlets}. These contain information regarding which subset of the data is taken from the
source, where it will be indexed in the destination, and what is the movement volume, all represented
by symbolic or constant expressions.
 
The simplest example is an assignment where no operands have side effects and no operands are function calls. In this case, we create a tasklet in a new state that contains the C code of the assignment. We augment the state by adding the data accesses as input and output memlets. For the example in \autoref{fig:sdfg}, the tasklet will have one outgoing memlet to the \texttt{y} array, and two incoming memlets from arrays \texttt{A} and \texttt{x}.

In the case of more complex assignments, we first identify sub-expressions (such as function
calls) with and without side effects and extract them. We create new assignments to temporary
values, which we replace in the C AST, maintaining the original evaluation order. 
The assignment will therefore be separated into multiple simple assignments, each analyzed separately, creating a semantically equivalent code that can be transformed into an SDFG, as seen in Figure~\ref{fig:sdfg} where the indirect array access \texttt{y[i]=A[col\_idx[i]]*x[j]} becomes \texttt{int idx=col\_idx[i]; y[i]=A[idx]*x[j];}.
This process is repeated recursively until a set of simple assignments is created. This effectively brings array indices into a static single assignment (SSA) form. This ensures individual array indices cannot change within a single SDFG state. 


\subsection{Statements}
\label{sec:statement}

In SDFGs, states are connected by inter-state edges that can have conditions or assignments (as
symbolic expressions) attached to them, controlling state transitions. This allows us to represent all
control flow constructs from C. An example \texttt{for} loop is shown in Figure~\ref{fig:sdfg}. 

%
%

\subsection{Functions}
\label{sec:func}

We differentiate between the functions whose source code we can access and external functions or library calls. 

\begin{figure}[h]
\begin{center}
    \vspace{-0.6em}
    \includegraphics[width=.9\linewidth,page=1,clip,trim={0cm 9cm 8cm 0cm}]{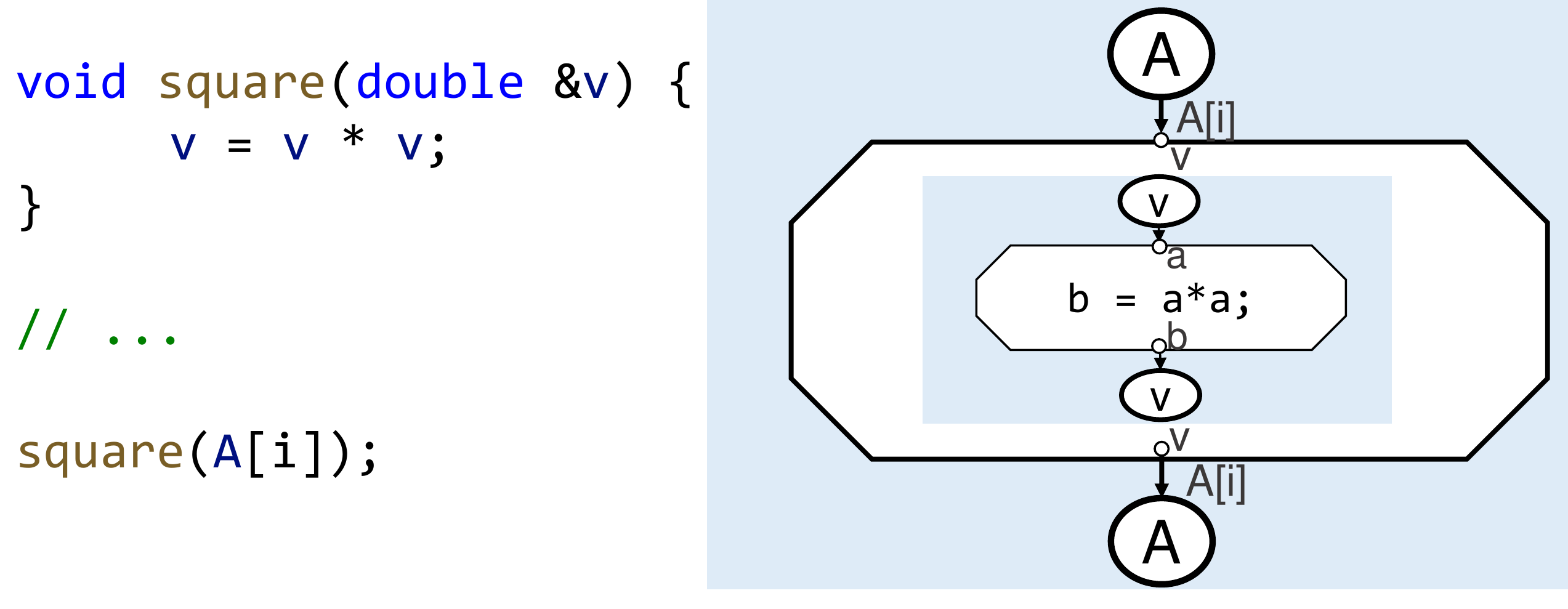}
    \caption{Function example.}
    \label{fig:func_example}
\end{center}
\vspace{-1em}
\end{figure}

\paragraph{Function calls with source} For these functions we create a nested SDFG and continue the analysis by creating a new context. When generating a new context for a function call, we prune unused parameters by taking the intersection between the union of arguments and global variables, and the union of the memlets of the
nested SDFG. An important aspect is that when changing the context through a function call,
it is possible for the view of data containers to change. For example, just one row of a two dimensional array can be passed as an argument. We track such behavior, as it can expose additional optimization opportunities. For example, the squaring function in Figure~\ref{fig:func_example} operates on a single value of $A$.

\paragraph{External and library calls}
\begin{wrapfigure}{r}{0.4\linewidth}
   \vspace{-1.25em}
   \begin{center}
     \includegraphics[width=\linewidth,page=3]{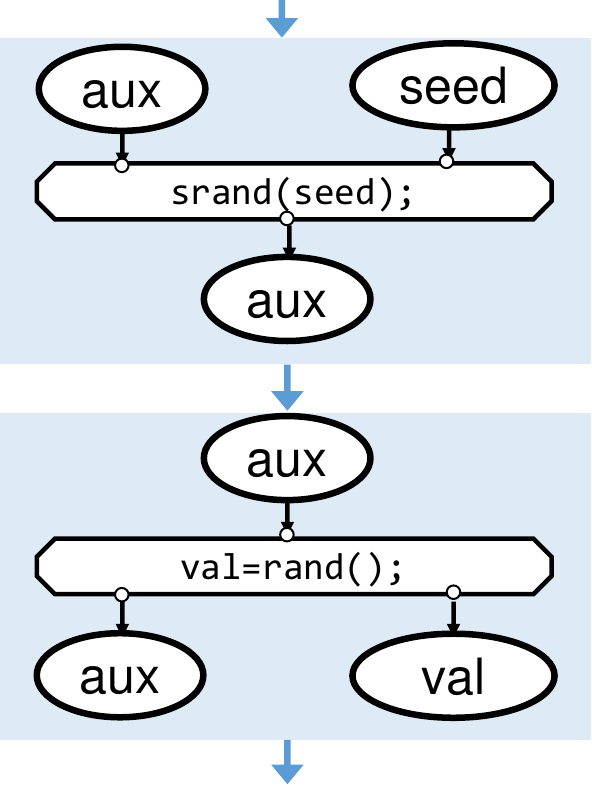}
     \caption{Stateful library call example.}
    \vspace{-0.5em}
    \label{fig:lib_example}
   \end{center}
   \vspace{-1.0em}
\end{wrapfigure}
\label{sec:external} Without any information about what happens within these calls, we create a tasklet containing the function call and assume all data containers accessible are read,
and all those that can be written, are.
We  define a list of libraries and functions that are stateless.
In all other cases, such as the \texttt{random} family of functions in \texttt{stdlib} or MPI, we define a unique global auxiliary scalar value used to augment the tasklet with, and additional input and output memlet to this scalar.
Given that this is the only use of this scalar, it allows dataflow optimizations to be applied to the rest of the application while ensuring that all calls to a stateful library are executed in program order. If no assumption can be made about which stateful libraries a particular function call might affect, we must assume it affects all.
In the example in Figure~\ref{fig:lib_example} we see how this addition ensures that the call to \texttt{rand()} cannot be reordered before \texttt{srand()}. Without the auxiliary variable, the two states would have no dataflow connecting them and their order might be incorrectly swapped.

\subsection{Optimization barriers} 

To be transformed correctly by our approach, the input code must not exhibit
undefined behavior (UB). Existing tools for detecting UB can be leveraged~\cite{hathhorn2015defining} in order to reject programs not meeting this restriction, and most applications should not exhibit UB.  One such example is not allowing pointer arithmetic with pointers not pointing to the same data container~\cite[Sec.~6.5.6]{csem}.

Furthermore, function pointers are not allowed, as they introduce a significant source of uncertainty when analyzing dataflow from source code and are not supported by our approach. Recursion and \texttt{longjmp}s are also not currently supported.

Finally, we differentiate between full program analysis and optimizing a subset of program such as a function call. If we are analyzing a function with multiple pointers as input we must conservatively assume such data containers might alias - significantly limiting the optimizations applicable. Analyzing the entire application rather than a function ensures all pointers and allocations are accounted for and overcomes this issue.

\subsection{Towards parallelism} \label{sec:parallelism}
A central property that allows SDFGs to achieve high performance is their inherent representation of parallelism.  
Since many parallel applications are composed of repeating subunits, a \textbf{Map} scope represents a parametrically-replicated subgraph. Maps have a symbolic integer set of variables, representing the range of values to replicate over. 
They are designed to be executed in parallel, and in the DaCe framework they can be used to enable various optimizations (such as vectorization, double-buffering), and generate multicore CPU code, GPU kernels (with static or dynamic load balancing), or FPGA programs (with support for pipelined and parallel components) --- once a Map is present in an SDFG, it can be optimized to state-of-the-art performance.
 Therefore, all transformations we create from this point on are in the service of this goal.

\section{Dataflow Extraction and Coarsening}
\label{sec:extraction}
Translating C to a semantically-equivalent SDFG yields states in program order with fine-grained control flow structure. In order to unlock the potential of dataflow analysis we lift dataflow by coarsening the control flow structure of our program and allow symbolic analysis of data access patterns.
%

%

Towards this goal, we apply three passes in a repeating fashion, the first two being novel contributions of this work: 
\begin{itemize} 
    \item \textbf{Symbolic scalar analysis}: Converts scalar variables to symbols. Assignments and conditions become state transitions rather than tasklets.
    \item \textbf{Access pattern propagation}: Computes the number of executions and symbol values in each state by propagating information regarding access patterns through state transitions, and allows automatic, static, work-depth analysis.
    \item \textbf{Dataflow Transformations}: We run a set of graph rewriting transformations that neither modify the program's result nor harm performance, which we call \textit{dataflow coarsening}.
    These transformations were introduced by Ben-Nun et al.~\cite{dace} and modify dataflow by merging states and nested SDFGs, and removing redundant data movement.
\end{itemize}
In the rest of the section, we elaborate on the algorithms and their computational complexity.

\subsection{Symbolic scalar analysis}\label{sec:scal2sym}

\paragraph{Motivation.} Optimizing data movement requires a detailed understanding of data access patterns: both which data structures are involved in each computation and which elements are accessed. Accesses that depend on data values cannot be known statically, but can be expressed symbolically --- either exactly, or through a conservative approximation. We will transform the program representation to create as many opportunities for symbolic analysis as possible. 

\paragraph{Approach.} In the translated C SDFG, every declaration is mapped to a data container, and every subsequent access is assumed to be a data-dependent access. As a result, all array accesses become indirect (i.e., requiring two round-trips to main memory), which inhibits further analysis. 
However, certain size-1 arrays and scalars, including all array access indices (as per Sec~\ref{sec:assign}) fulfill the conditions of SDFG symbols (\S~\ref{sec:declare}), namely that their value does not change throughout the course of a state, and their initialization and updates can be expressed as symbolic expressions. In such cases, those scalars become symbols that are set in inter-state edges.

This replaces tasklets with symbolic control flow, it enables both structured control flow detection and narrows memlet accesses to symbolic sets.
With the former, it opens up the opportunity to perform symbolic range analysis for regular control flow constructs such as loops and branches (used, e.g., for Work-Depth program analysis, \S~\ref{sec:wd}).
With the latter, the symbolic memlets in turn enable (1) dependency analysis and parallelism extraction in various granularities;
(2) potential data race detection when memlet access sets intersect and (3) the information
necessary to apply subsequent transformations, such as local storage/access deduplication (common subexpressions), vectorization (contiguity), inferring disjoint regions for distributed computing, and others. 

\begin{figure}[t]
    \centering
    C code: \texttt{int x = 5; /*...*/ B[x] = A[x] * A[x];}\\\vspace{0.5em}
    \includegraphics[width=.9\linewidth,page=4]{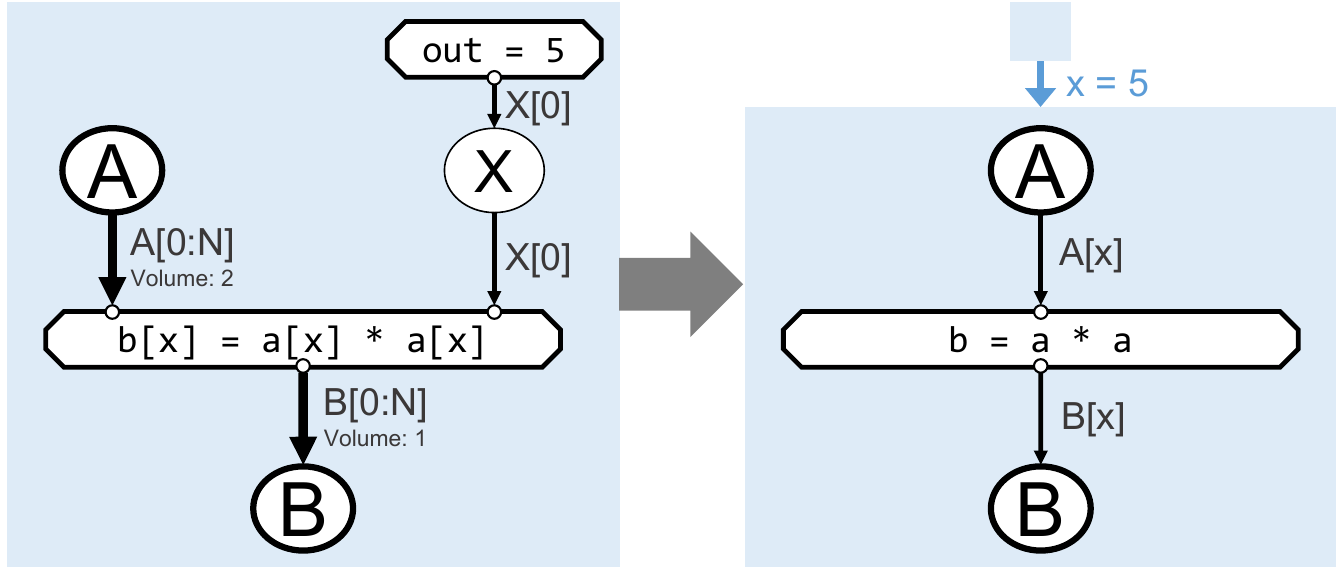}
    \caption{Symbolic scalar analysis example}
    \label{fig:scal2sym}
\end{figure}

In Figure~\ref{fig:scal2sym} shows a simple example of a scalar converted to a symbol. On the right side, it becomes clear that the access volume to \texttt{A} is 1 rather than 2, and that the access sets are the same. If this state machine is contained within a loop, it could be subsequently converted to a map.

A further example can be seen in Figure~\ref{fig:sdfg}. At the beginning of the analysis \texttt{idx} is a scalar. Our symbolic scalar analysis detects that \texttt{idx} can be transformed into a symbol and splits it out of the state into a new state, placing the assignment in an inter-state edge.

This is a crucial contribution and combined with the inter-procedural analysis allows optimization opportunities no other tool can detect - including the discovery of indirect access patterns.

\paragraph{Complexity} The algorithm is linear in the overall number of dataflow nodes $n$ and memlets in the SDFG states and all its nested SDFGs (denoted by $n_i$ and $m_i$ for SDFG $i$), as well as the inter-state edges $M$ of the top-level SDFG. The overall runtime is $O(M+\sum_i(n_i+m_i))$.

\subsection{Access pattern propagation}\label{sec:sprop}

\paragraph{Motivation.} Data movement optimizations are not found only in limited scopes such as within a loop, but across the entire program.  Therefore, analyzing program fragments separately does not suffice, and a holistic approach is necessary. For example, if we want to generate the access set of an entire function, we would need to know the overall potential subsets that each part may access, which may depend on nested loops or even the contents of the data containers. We introduce a method of propagating data movement across scopes and contexts. 

\paragraph{Approach.} In the case of SDFGs specifically, the problem lies in propagating data movement from Map scopes and arbitrary state machines inside nested SDFGs to input/output memlets in the outer state.

In the SDFG representation, memlets outside maps are inferred directly from the internal memlets in a process called \textit{memlet propagation}. This process computes the image set of the union of internal memlet subsets, when the map range is applied on them. Propagated memlet volume is the product of the map range size with the sum of volumes in each internal memlet. For example, a memlet \texttt{A[2*i+5]} propagated over a map ranged $i\in[0,N), N\in\mathbb{N}$ results in \texttt{A[5:2*N + 4:2]} (in Python index notation) with a volume of $N$.

%
To produce outer memlets, we can boil down the requirements further into two values for each state: number of \textit{executions}, and its corresponding \textit{symbol ranges}.
Loop counting (a subset of this process) is a known problem in program analysis~\cite{benamram13} and undecidable in the general case (due to the Halting problem). We thus turn to recognize certain structured control flow patterns and try to provide upper bounds otherwise.

\begin{figure}[t]
    \centering
    \includegraphics[width=\linewidth,page=5]{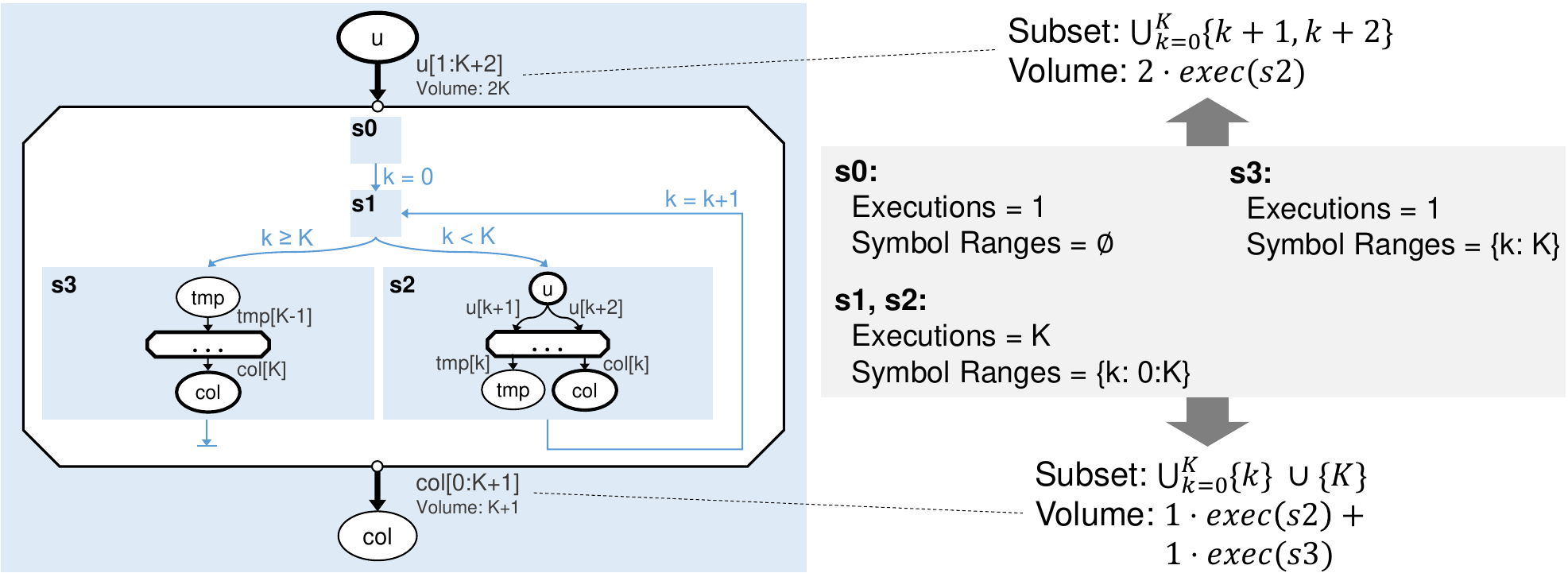}
    \caption{Access pattern propagation on a vertical stencil.}
    \label{fig:sprop}
\end{figure}

Access pattern propagation is another contribution of this work and operates bottom-up on the SDFG ``tree'' (where descendants of an SDFG are its nested SDFGs). Following memlet propagation, we begin by detecting well-structured loops (i.e., iterating over symbolic ranges) and branches using standard CFG techniques. For the former, we employ cycle detection, annotating loop guards and symbolic loop ranges on loop bodies. Unrecognized back-edges annotate the destination state with an \textit{unbounded} number of executions. For the latter, we compute the dominance frontiers for the state machine (ignoring back-edges) to identify branch merge states.
Then, we perform a modified DFS traversal of the state machine to accumulate state executions and propagate them forward, keeping a traversal state to pass along through outgoing inter-state edges. We use the following rule-set: 
\begin{itemize}
    \item Start state is executed once and without symbol ranges. 
    \item If a state has one outgoing edge, the same number of executions and symbol ranges propagate directly, where if a symbol is assigned a value, its union with the current range is computed. 
    \item In case of branching, each branch is annotated as ``bounded by'' the number of executions (since each branch state may be entered up to the total number of tests). For branch merge states, the executions change back to exact. 
    \item Loops are annotated with symbolic execution using all available ranges from the annotation phase (we use nested sums in the SymPy symbolic math engine).
    \item Unbounded states propagate forward as unbounded.
\end{itemize}
Outer memlets on the nested SDFG follow similar rules as memlet propagation: access subset is the image set of the union of memlet subsets, applied to the state symbol ranges; outer volume is the sum of memlet volumes multiplied by their respective state executions.
The process is then repeated on the parent SDFG after its memlet propagation, until the root is reached.
An example of the process is shown in Figure~\ref{fig:sprop}, on an excerpt of a vertical stencil used in numerical weather prediction models.
In this case, only after state propagation can we apply, for example, the \texttt{MapFusion} transformation for fusing GPU kernels and storing \texttt{col} on local registers rather than global memory.
%

\paragraph{Complexity}
For a tree of $S$ SDFGs, with up to $N$ states and $M$ inter-state edges, the complexity of access pattern propagation is $O(S\cdot (N+M+DF(N, M)))$. Both cycle enumeration and our custom depth-first traversal take $O(N+M)$, and $DF(N, M)$ is the complexity of computing dominance frontiers~\cite{cooper01}.

\section{Dataflow Optimization}\label{sec:optimizing}

Once the dataflow is coarsened, many opportunities open up for parallelizing and optimizing the input application.
In particular, since 
(a) memory can be traced for aliasing (\S~\ref{sec:declare});
(b) we are guaranteed that external side effects are visible through per-library state (\S~\ref{sec:func});
(c) all the loop ranges are symbolic values that can be reasoned about (\S~\ref{sec:scal2sym});
(d) all memlets have been propagated outside of all nested SDFGs and Maps (\S~\ref{sec:sprop}); loop parallelism feasibility becomes straightforward. All that is necessary to perform is a check of the memlet subsets in the loop body states.

\subsection{Detecting updates}

To open more opportunities for parallelization, we detect and create a specialized abstraction for associative operation updates.
\textit{Write conflicts} refer to the situation where two data movements end in the same location in parallel. Update operations are functions that receive the old and new value and perform an update atomically.

This is a stepping stone and allows the efficient automatic parallelization of loops containing specific read-modify-write patterns, as we shall show.

\subsection{Autoparallelization (Loop to Map)}
\label{sec:l2m}

Tying all contributions together, we present a method to transform serial loops into parallel parametric maps using static analysis of SDFGs.

A loop over the iteration space $I$ can be turned into a parametrically parallel Map scope if it creates no data races once parallelized. Formally, for each data container with the read and write access sets $r_i$ and $w_i$ in iteration $i\in I$, we need to guarantee two conditions for every pair of iterations $i,j\in I$, where $i\ne j$: $w_i \cap w_j=\emptyset$ and $r_i \cap w_j=\emptyset$. Finding those intersections for certain classes of loop ranges and access sets (e.g., affine expressions, polytopes) can be done with existing tools (e.g., \texttt{isl}~\cite{isl}), and we opt to do similarly with the SymPy library and set intersection, and leverage the symbolic access information provided by the data-centric view of the program.

While this covers ``embarrassingly-parallel'' loops, loops like the one below include reductions:

\begin{lstlisting}[language=C]
int i,j,k;
for (i = 0; i < N; i++)
  for (j = 0; j < M; j++)
    for (k = 0; k < K; k++)
      C[i][j] = C[i][j] + A[i][k] * B[k][j];
\end{lstlisting}

Such loops do not qualify for the two conditions, as for any triplet of \texttt{i,j,k} we see that \texttt{k} does not participate in the write access set of $C$. However, notice that the read value of \texttt{C[i][j]} is not used apart from the modification part. In our parallelization pipeline, we convert such compound assignments to update-memlets (via a graph-rewriting transformation). This removes the read access, replacing it with a single outgoing update-memlet. For any update-memlet in a data container access set, we relax the write-write conditions (but not read-write) to ignore it, thus enabling such cases to become parallel.
Our parallelization approach assumes associative reductions, and reverts to serialization without this assumption.

There are other benefits to this update analysis abstraction, for example when indirect memory access is used in writes. Since we symbolically analyze the memlets, we can detect that the same element, albeit known at runtime, is modified, and convert it. This allows us to analyze codes such as the following example, taken from the LULESH scientific application, \S~\ref{sec:lulesh}:

\begin{lstlisting}[language=C]
for (Index_t lnode=0; lnode<lnodes; ++lnode) {
  Index_t gnode = elemToNode[lnode];
  domfx[gnode] += fx_local[lnode];
  domfy[gnode] += fy_local[lnode];
  domfz[gnode] += fz_local[lnode];
}
\end{lstlisting}

While automatically-detected parallel sections are powerful, they are not sufficient as-is to optimally utilize high-performance hardware (apart from mapping into multiple CPU cores - for example by generating OpenMP loops and using atomic reductions to avoid data races).
At this stage, though, we can mutate the analyzed dataflow by modifying the computation schedule, introducing temporary buffers, and changing data types, all without changing program semantics as shown by Ben-Nun et al~\cite{dace}.

\section{Evaluation}\label{sec:eval}
To evaluate our approach, we consider both the degree to which we expose parallelism, and the performance of code we generate following the pipeline in Sections \ref{sec:extraction}--\ref{sec:optimizing}.
We evaluate the 30 tests included in the Polybench~\cite{polybench} suite, and the LULESH~\cite{lulesh} unstructured grid scientific application.

\subsection{Work-Depth analysis}\label{sec:wd}

As an objective means of measuring the effectiveness of our approach towards exposing parallelism, we use the \textit{Work and Depth} model to compare the different stages of the process.

\begin{figure*}[t]
    \centering
    \subfloat
	{
		\includegraphics[width=1.0 \textwidth]
		{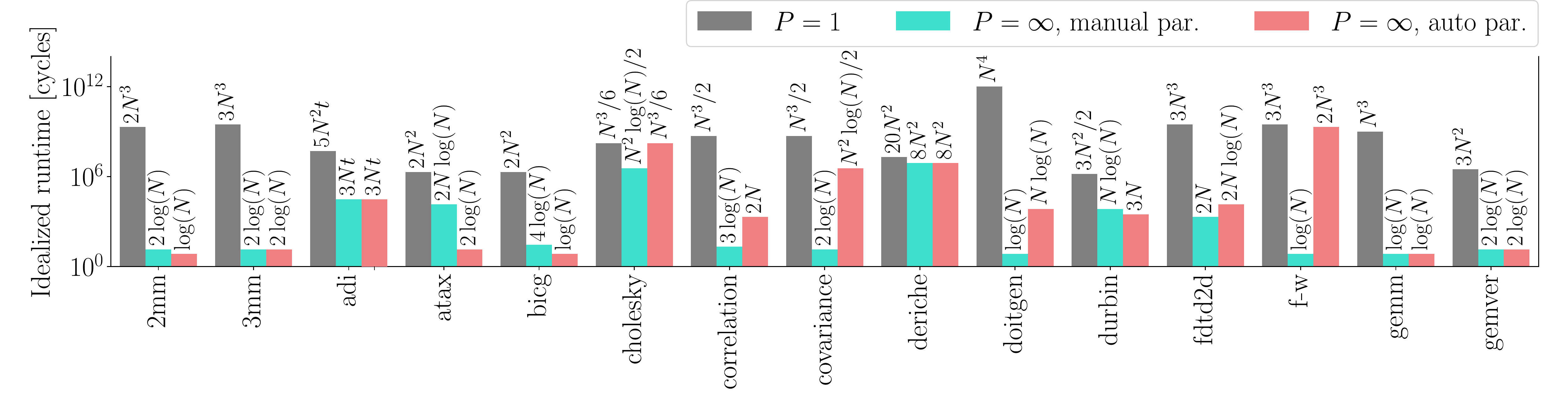}}
	\\
	\subfloat
	{
		\includegraphics[width=1.0 \textwidth]
		{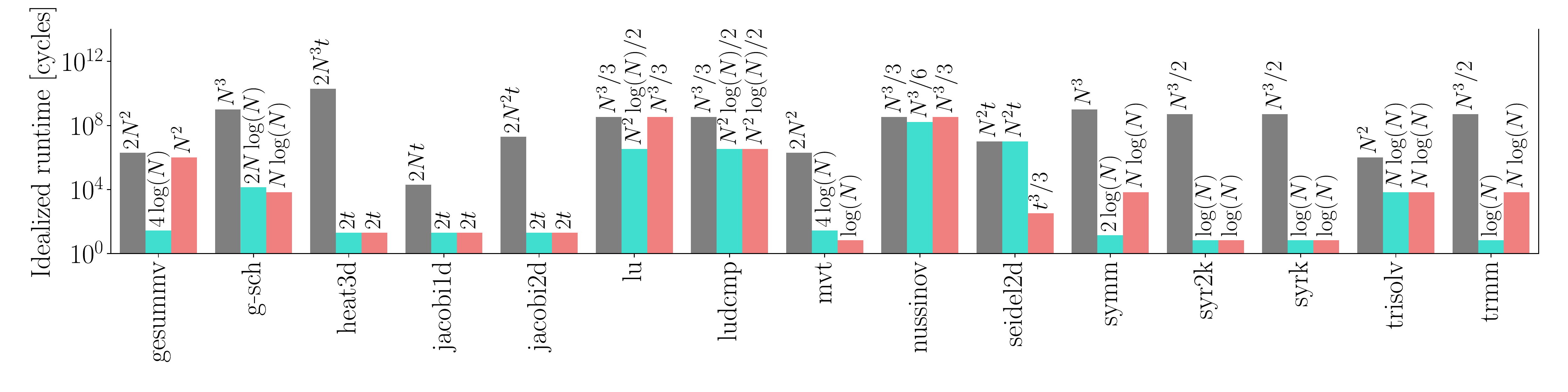}}
    \vspace{-1.5em}
    \caption{Theoretical time required to execute Polybench kernels --- hand-tuned vs. automatic parallelization. Sample evaluation for all array sizes $N=1,000$, time steps $t=10$ (where available).}
    \label{fig:depth}
\end{figure*}

Briefly, the model states that any computation can be represented by a DAG (called a computational DAG, or CDAG). The nodes represent computations whereas the edges represent dependencies. We can then characterize the computation by the number of nodes $W$ (\textit{Work}) and the longest path on the CDAG $D$ (\textit{Depth}). With these two parameters we can evaluate, for example, the average parallelism in an application, by taking $W/D$, which is the average number of concurrent nodes in any level of the CDAG. In Figure~\ref{fig:depth}, we list the recovered parallelism on Polybench by computing the asymptotic number of cycles on one and an infinite number of processors $P$ for our work and hand-tuned~\cite{dace} versions.

\paragraph{Deriving work and depth}
We can directly derive the work and depth of a computation statically from a fully-analyzed SDFG.
Observe that the only sources of computational work in SDFGs are tasklets.
Given our construction of tasklets from C expressions, we assume that each tasklet evaluation takes one unit.
Therefore, if a tasklet is in a Map scope of size $n$, the induced work is $n$ with depth $1$. If it is in a sequential loop, the corresponding depth is also $n$. To minimize depth, associative reductions in a CDAG can be represented by a binary tree. Thus, any write-conflict resolution on a parallel scope incurs at least a depth of $\log r$, where $r$ is the size of reduction dimension. For example, in matrix multiplication a tasklet exists in a 3-dimensional Map sized $NMK$, with reduction dimension sized $K$. Therefore, induced work is $W = MNK$, and depth is $D = \log K$.

Work and depth are propagated through states and recursively upwards in the SDFG tree. Total work is the sum of all the tasklets' and nested SDFGs' work, and total depth is evaluated by finding the longest path, weighted by depths of nested SDFGs and number of state executions.

In Figure~\ref{fig:depth}, the results show that most applications (25/30) are able to improve upon the $P=1$ case, detecting parallelism automatically. While the hand-tuned version does contain more Map scopes, in many of the cases the depth matches exactly or is close, and in two cases (\texttt{atax}, \texttt{bicg}) the automatic parallelization scheme even detected missed opportunities in the hand-tuned code. The full work and depth formulas are available in the supplementary material.

\subsection{Case studies}\label{sec:casestudies}

We measure performance on a dual-socket $2{\times}18$ core Intel Xeon Gold 6154 system clocked at $3.00$~GHz with 384 GB RAM. SDFGs were compiled using \texttt{gcc}~10.2.0, and compared with \texttt{gcc}, Clang~11.0.0 with Polly~\cite{polly}, Pluto~0.11.4~\cite{pluto}, and \texttt{icc}~19.0.5.281. We verify all results  within $10^{-5}$ absolute tolerance, and report median performance of 10 runs with error bars representing 95\% confidence intervals. We use the \texttt{-ffast-math} compiler flag on all benchmarks except \texttt{gramschmidt}. The \texttt{gramschmidt} benchmark is not numerically stable and would produce incorrect results if the \texttt{-ffast-math} compiler flag is used. We do not tune tile sizes neither in SDFGs nor Pluto.

\subsubsection{SpMV} The dataflow coarsening and optimization of SpMV leverages all concepts we have introduced: symbolic scalar analysis to understand indirect access patterns, update detection to allow the nested loops to be parallelized, and finally the automatic parallelization of the nested loops.

The parallel code generated by our approach uses OpenMP parallel loops and, where necessary, atomic accesses to ensure no data races occur. If desired, the code could further be optimized by applying tiling or other transformations, or by manually improving the resulting source code.

Our automatic workflow leads to parallel code with a runtime on par with that achieved by \texttt{icc -parallel} and approximately 6 times faster than the best results among other compilers and autoparallelizing tools, as seen below: 

\begin{center}
{
\vspace{0.2em}
\setlength{\tabcolsep}{0.3em}
\begin{tabular}{ r r r r r r r}
  DaCe & polly & pluto & icc -parallel & icc & gcc & clang \\
 \hline
 0.54s & 3.95s & failed & 0.45s & 4.98s &3.55s & 4.19s \\ 
 \hline
\end{tabular}
}
\vspace{0.2em}
\end{center}

This is explained by polyhedral compilers being unable to identify and expose parallelism in this case. From the tools and compilers we investigated, only \texttt{icc} managed to create a parallel version of the code.

\subsubsection{Polybench}

\begin{figure*}[t]
    \centering
    \includegraphics[width=\linewidth]{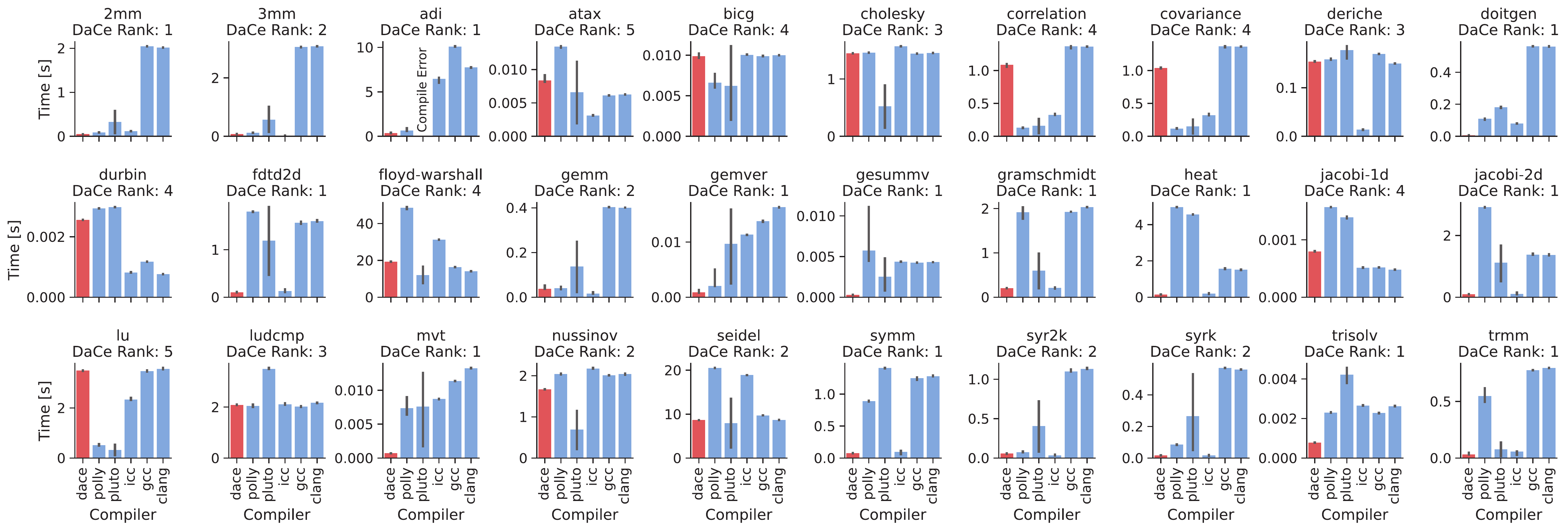}
    \vspace{-1em}
   

    \caption{Polybench kernel performance comparison against state of the art compilers and autoparallelizing tools.}
     \label{fig:polybench-kernel}
\end{figure*}

We use DaCe to transform the C code of the entire tests (not just the kernel) to SDFGs, in order to capture the full application dataflow. The entire SDFG processing pipeline takes 1--7 seconds (3.7 on average) for a Polybench application in our Python-based framework. This is comparable with Pluto, which runs for 0.12--24.69 seconds (1.62 on average) on the same machine.

In Figure~\ref{fig:polybench-kernel}, we evaluate the performance of our parallelized applications with the compilers and their respective auto-parallelizers (icc-parallel, polly-parallel, Pluto parallel/lbtile/multipar), taking the flags and configurations that yield the fastest runtimes. We compare the runtime of the kernels themselves as per the polybench benchmark. Our approach outperforms all others on 13 of 30 tests, making it a powerful tool to ensure unmodified C applications can run efficiently in parallel. For 9 other tests, our approach is among the top 3 best performing options, while for the last 8 we are as fast as state of the art C compilers.

The data-centric view of programs allows us to optimize and parallelize where other tools cannot, such as in the case of \texttt{gemver}, \texttt{gesummv}, \texttt{mvt}, and \texttt{trisolv}. Polyhedral analysis is orthogonal to the dataflow representation, and can find other opportunities for parallelization in some cases, such as correlation. In the next step, a performance engineer could improve performance by invoking additional SDFGs transformations.
In some cases such as \texttt{nussinov} and \texttt{floyd-warshall}, the code is structured around control flow rather than dataflow, preventing most of our optimizations. Even in such cases, performance does not degrade - we are simply on par with C compilers such as gcc or icc.

In summary, our data-centric approach outperforms each other compiler and auto-parallelizer on the majority of Polybench tests, remains competitive on all others, and does so without needing code annotations or manual guidance.

\subsubsection{LULESH} \label{sec:lulesh}

\begin{figure}[t]
    \centering
    \includegraphics[width=\linewidth]{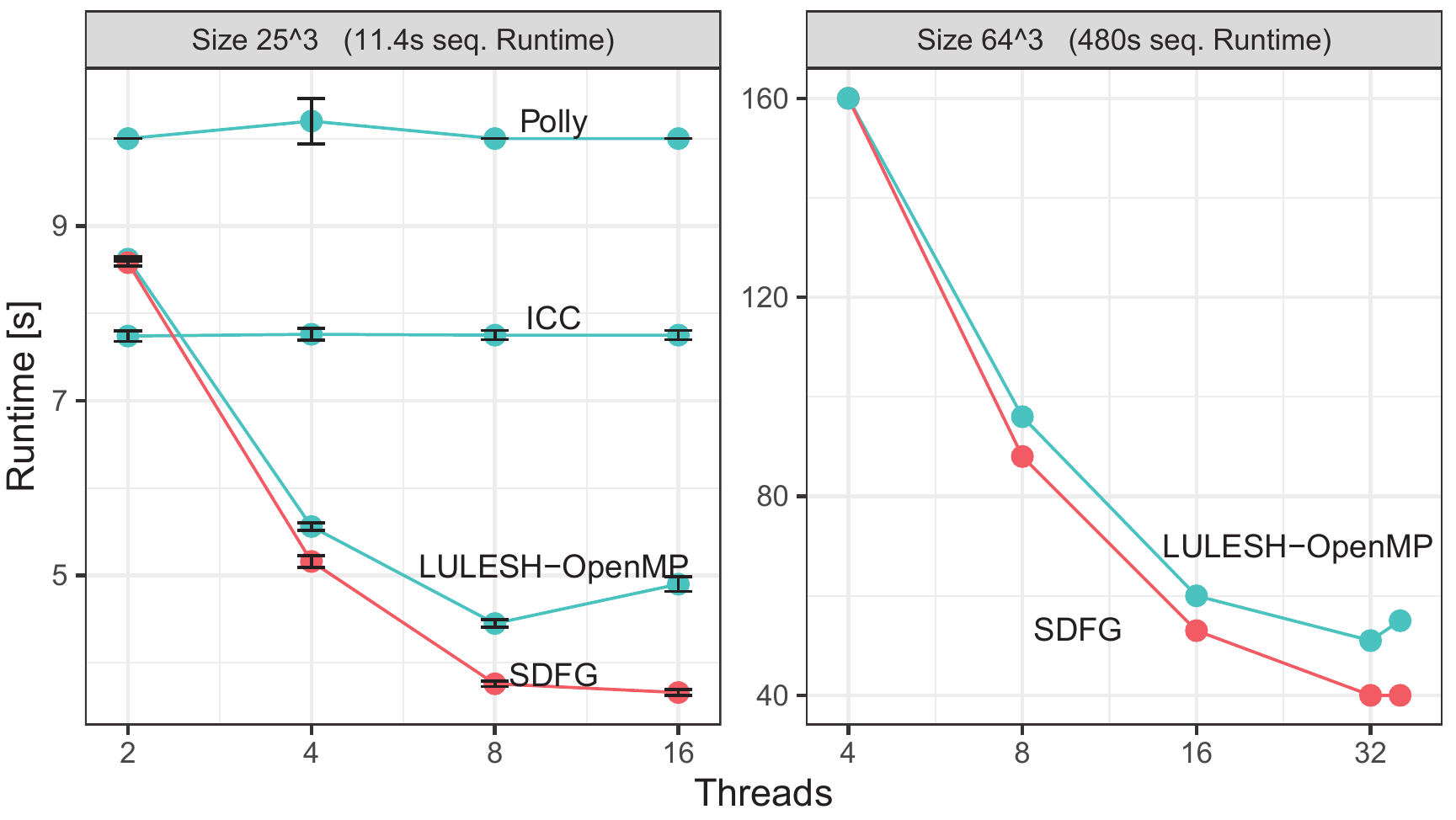}
    \vspace{-2em}
    \caption{LULESH parallel performance comparison.}\vspace{-1em}
    \label{fig:evallulesh}
\end{figure}

The Livermore Unstructured Lagrangian Explicit Shock Hydrodynamics (LULESH) application is an unstructured physics simulation. It is written in C++ rather than C, but like many HPC applications only uses a few C++ features, encapsulating data in a class object. 
We only modify LULESH to allow C-to-SDFG processing by replacing this monolithic data structure with its components. 

LULESH has been extensively optimized and parallelized to serve as a proxy-app for Exascale co-design. Towards this goal, its developers implement a complex and extensive shared memory parallelization scheme in OpenMP using multiple additional data structures, as well as thread-local storage to minimize communication and sharing when accumulating results. To show the power of data-centric analysis, we only consider the sequential code, disregarding the user directives and shared memory scheme. 
We do however compare the performance of our approach to this manually-tuned shared memory parallelization (LULESH-OpenMP). 

We focus our analysis on the \texttt{CalcVolumeForElems} function and other functions it calls. This amounts to 895 lines of code, contributing over 60\% of the total application runtime, and the largest contiguous code region executed between distributed communication steps using MPI.

Since LULESH operates on unstructured grids, most of its data access patterns are indirect, which prevents automatic parallelization tools from identifying most loops as being parallelizable. Both \texttt{icc} and Pluto~\cite{pluto} found no parallelizable loops within the analyzed scope (the latter due to missing required annotations), and Polly-parallel~\cite{polly} found two loops: one trivial loop in the which three arrays are set to zero and one loop with a range of $[0,4)$. 

Through the aforementioned data-centric transformations, \textit{our approach correctly detects that \textbf{all} 16 loops within the scope of our analysis can be parallelized}, 5 of those iterating over the entire problem size --- and complete the entire analysis pipeline in 51 seconds. 

We compare our approach with the manually tuned version provided by the developers for both sequential and parallel executions in Figure~\ref{fig:evallulesh}. The data-centric view of the entire program coupled with symbolic analysis of array access patterns allows indirect memory to be analyzed in a way not available to other techniques. The figure empirically verifies the missed parallelism opportunities by \texttt{icc} and Polly, and shows that the SDFG of the simple sequential code outperforms the original implementation by up to \textbf{21\%}. Upon inspection of the code, the speedup stems from more scalable memory management and reduction scheme of the parallelized SDFG over the manually-tuned counterpart.

\section{Discussion and Related Work}

Optimizing for data movement and locality is an evolving research topic~\cite{trends}.
All popular C compilers optimize data movement, and several (e.g., \texttt{icc}, \texttt{gcc}) attempt to automatically parallelize and vectorize input codes. There exist tools such as Polly~\cite{polly},  Pluto~\cite{pluto}, and DiscoPoP \cite{discopop} focused on discovering parallelism. They are, however, limited in their optimizations by not being able to track dataflow through indirection or across program scopes.

Language extensions focused on exposing parallelism, including OpenMP~\cite{openmp}, OpenACC~\cite{openacc}, and XScalableMP~\cite{xscalablemp}, require user annotations to function. While this allows C programs to efficiently use parallel hardware resources, extensive tuning is necessary to achieve this in practice.

SYCL~\cite{sycl}, a cross-platform abstraction layer, provides port-able performance by using standard C++ along with template and generic lambda functions to create source files containing code for multiple heterogeneous architectures. 
Lift~\cite{lift} is similar to SYCL in that programs are written in a high-level language but also provides primitives for expressing common parallel concepts such as map and reduce and provides rewrite rules to map these high-level programs to OpenCL.

Other intermediate representations (IR) are used for optimizing data movement and exposing parallelism in C.
In the LLVM IR~\cite{llvm} basic code blocks, transformed into Single Static Assignment~\cite{ssa91efficient, ssa88gvn} form  can be represented as a directed acyclic graph. However, the LLVM IR currently lacks analysis features necessary to track dataflow through different scopes and consequently coarsen dataflow as we do in SDFGs. The same applies to MLIR~\cite{lattner2020mlir}, with an important distinction: MLIR allows to define dialects of LLVM IR, with specific high-level transformations. The same concept is present in SDFGs in the form of library nodes~\cite{stencilflow}.

HPVM~\cite{hpvm} is a dataflow-graph based IR, similar to SDFG states. Instead of lowering to C, ``tasklets'' are expressed in LLVM IR instead of C, and coarse grained control flow is outside of the scope of HPVM. This is an important difference, as including both coarse-grained control flow and dataflow in SDFGs is necessary for the presented graph transformations.

Program Dependence Graphs (PDGs) \cite{pdg87} offer a control-centric representation in which statements and predicate expressions are given as nodes, and data dependencies and control flow are depicted with edges. While this model works well for load/store architectures (e.g., CPU, GPU), SDFGs with explicitly defined state machines of dataflow execution are better suited to reconfigurable hardware as well.
The SDFG also bears similarities to dataflow-based functional programming languages, such as Id~\cite{id}, VAL~\cite{val}, and SISAL~\cite{sisal}, but differs in the explicit representation of memory (allocation, addressing, in-place operations) and the parametric, statically-analyzable dataflow definition. Furthermore, the state machine around the dataflow in SDFGs, which introduces procedural constructs, is easier to convert from C than purely functional languages.

Many approaches perform their data-locality optimizations within the polyhedral model~\cite{bondhugula2008automatic,sbirlea15,zinenko2017unified,pluto,reconfigurable}. This model is fundamentally different from SDFGs: every access must be (semi-)affine and iteration spaces are defined as polytopes. Affine transformations can be used to perform locality optimizations such as tiling. Polyhedral compilers are beneficial for such loops (finding nontrivial schedule optimizations as in Figure~\ref{fig:polybench-kernel}). SDFGs complement this by handling non-polyhedral codes and non-affine transformations.

Many systems~\cite{legion,kokkos,raja,gunrock,spiral} and representations~\cite{Halide,bamboo,isard2007dryad,murray2013naiad,pencil,chapel,gajinov12,pipes,spatial} perform dataflow optimization but do not have C frontends. 
However, they share important concepts with the SDFG representation.
Halide~\cite{Halide} provides a domain-specific framework for the optimization of stencil/image-processing kernels, which aims to make dataflow optimizations easier by decoupling computation from data layout and scheduling. Data-layout and scheduling changes are composable, similar to graph transformations in SDFGs. 
Legion~\cite{legion} is centered around managing dependencies between subtasks and scheduling them at runtime.
\section{Conclusion}
\vspace{-0.5em}

We present a workflow that lifts dataflow semantics out of C code. By using a data-centric IR, we expose optimization and parallelization opportunities normally unavailable to C developers and do so \emph{without annotations}: we detect parallelization opportunities in multiple examples that other automatic parallelization tools miss.
Our workflow is fully automatic, transparent, and fast, and allows us to outperform existing tools on various programs, including a developers' own parallel version of the LULESH simulation by up to 21\%.

With the power of other compiler analyses, such as devirtualization, the work could be extended to support a subset of C++, enabling automatic parallelization for the vast majority of performant codebases in the world. 

\begin{acks}
This work received EuroHPC-JU funding under grant no. 101034126, with support from the European Union’s Horizon2020 programme and from the European Research Council \includegraphics[height=1em]{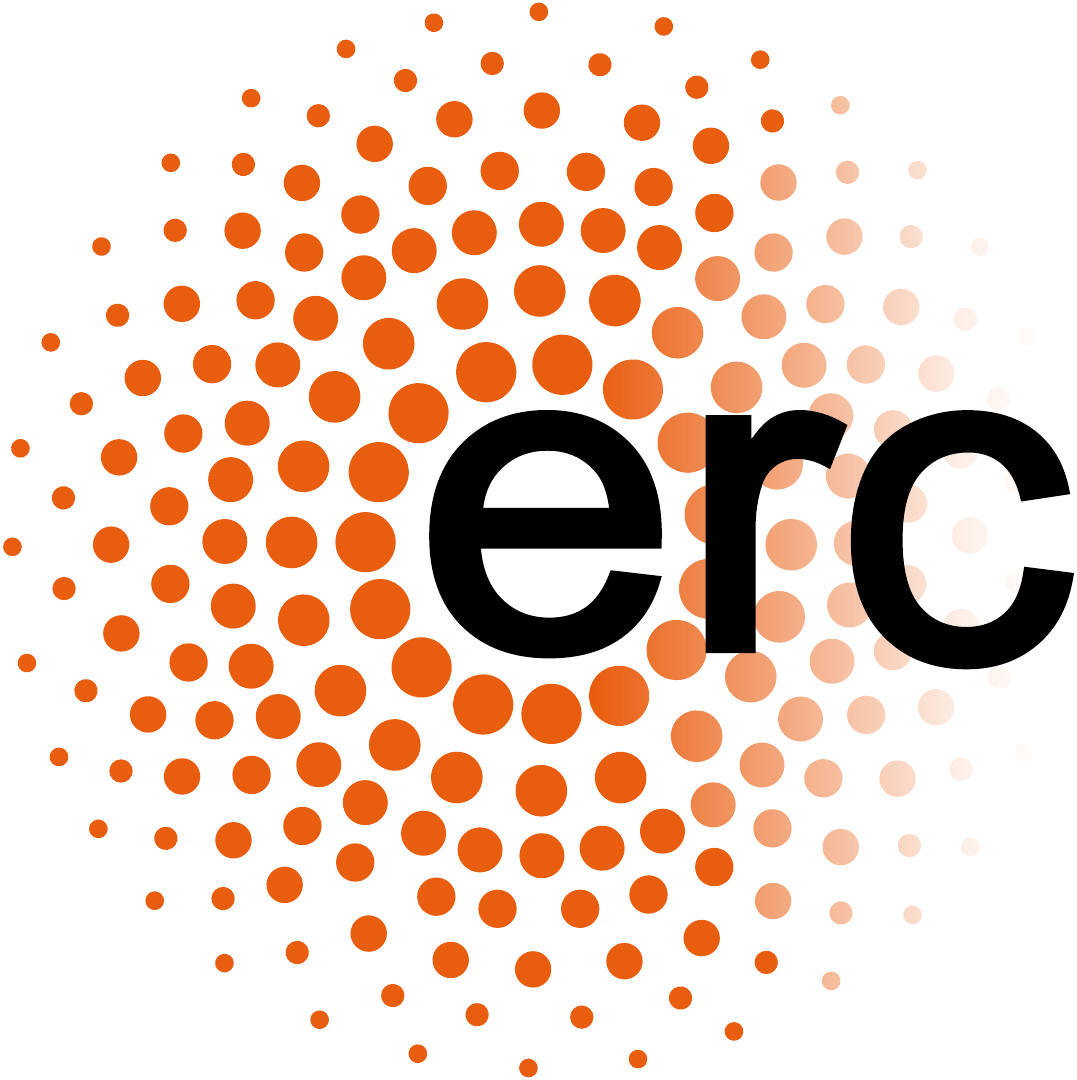} under the PSAP grant
agreement number 101002047. We also wish to acknowledge support of the RED-SEA and DEEP-SEA projects, grant numbers 955776 and 955606 respectively. We also wish to acknowledge the Swiss National Supercomputing Centre (CSCS) for
access and support of the computational resources.

\end{acks}

\bibliographystyle{ACM-Reference-Format}
\bibliography{references}

\end{document}


\title{Supplementary Material: C-Flow: Lifting C Semantics for Dataflow Optimization}

\author{Ben Trovato}
\authornote{Both authors contributed equally to this research.}
\email{trovato@corporation.com}
\orcid{1234-5678-9012}
\author{G.K.M. Tobin}
\authornotemark[1]
\email{webmaster@marysville-ohio.com}
\affiliation{%
  \institution{Institute for Clarity in Documentation}
  \streetaddress{P.O. Box 1212}
  \city{Dublin}
  \state{Ohio}
  \country{USA}
  \postcode{43017-6221}
}

\renewcommand{\shortauthors}{Calotoiu, et al.}




\maketitle
\begin{table*}[h!]
\footnotesize
\begin{tabular}{p{0.70in} p{1.1in} p{1.9in} p{1.3in}}
\toprule
 
  &	        & \multicolumn{2}{c}{\bf Depth}  \\
  \cmidrule{3-4}
 \bf Kernel       &	\bf Work            & \bf Hand-tuned                &\bf Auto-parallelized  \\
\midrule
2mm & $2N^2(N + 1)$ & $2\log(N) + 3$ & $5\log(N) + 4$\\ \addlinespace 
3mm & $3N^3$ & $2\log(N) + 2$ & $4\log(N) + 4$\\ \addlinespace 
atax & $N(2N + 1)$ & $N(\log(N) + 1) + N + 1$ & $2\log(N) + 4$\\ \addlinespace 
bicg & $N(N + 1)$ & $2\log(N) + 2$ & $\log(N) + 3$\\ \addlinespace 
cholesky & $N(N^2 + 3N + 2)/6$ & $N(N + (N - 1)(\log(N - 2) + 1) + 2\log(N - 1) + 3)/12$ & $N(N^2 - 2N + \log(N - 1) + 3)/6$\\ \addlinespace 
correlation & $N(N^2 + 6N + 5)/2$ & $3\log(N) + 7$ & $2N + (N - 1)^2\log(N) + 3(N - 1)^2 + \log(N) + 6$\\ \addlinespace 
covariance & $N(N^2 + 7N + 6)/2$ & $2\log(N) + 7$ & $N^2\log(N) + 4N^2 + N + \log(N) + 3$\\ \addlinespace 
deriche & $2N(3N + 2)$ & $4N^2 + 4N + 2$ & $2N(3N + 2)$\\ \addlinespace 
doitgen & $N^3(N + 2)$ & $\log(N) + 3$ & $N(2\log(N) + 3)$\\ \addlinespace 
durbin & $3N^2/2 + 3N/2 - 2$ & $4N + (N - 1)(\log(N - 1) + 1) - 3$ & $4N + (N - 1)(\log(N - 1) + 1) - 3$\\ \addlinespace 
fdtd2d & $N(3N^2 - 3N + 1)$ & $3N$ & $2N$\\ \addlinespace 
floyd-warshall & $N^3$ & $\log(N) + 1$ & $N^3$\\ \addlinespace 
gemm & $N^2(N + 1)$ & $\log(N) + 2$ & $2\log(N) + 2$\\ \addlinespace 
gemver & $N(3N + 1)$ & $2\log(N) + 4$ & $2\log(N) + 4$\\ \addlinespace 
gesummv & $N(N + 1)$ & $2\log(N) + 2$ & $N + 2$\\ \addlinespace 
gramschmidt & $N(2N^2 + 3N + 3)/2$ & $N(2\log(N) + 7)$ & $N(2\log(N) + 7)$\\ \addlinespace 
heat3d & $2t(N - 2)^3$ & $2t$ & $2t$\\ \addlinespace 
jacobi1d & $2t(N - 2)$ & $2t$ & $2t$\\ \addlinespace 
jacobi2d & $2t(N - 2)^2$ & $2t$ & $2t$\\ \addlinespace 
lu & $N(N^2 - 1)/3$ & $N(N + (N - 1)(\log(N - 2) + 1) + (N + 1)(\log(N - 1) + 1) - 1)/6$ & $N^2(N - 1)/3$\\ \addlinespace 
ludcmp & $N(2N^2 + 15N + 19)/6$ & $N(N\log(N - 2) + N\log(N - 1) + 6N - \log(N - 2) + 5\log(N - 1) + 12)/6$ & $N(3N + (N - 1)(\log(N - 2) + 1) + 3\log(N - 1) + 5)/3$\\ \addlinespace 
mvt & $N^2$ & $2\log(N) + 1$ & $\log(N) + 1$\\ \addlinespace 
nussinov & $N(N^2 + 9N - 10)/6$ & $N(N^2 - 3N + 2)/6$ & $N(N^2 + 9N - 10)/6$\\ \addlinespace 
seidel2d & $t(N - 2)^2$ & $t(N - 2)^2$ & $t(N - 2)^2$\\ \addlinespace 
symm & $N^2(N + 3)/2$ & $\log(N) + \log(N - 1) + 3$ & $3N$\\ \addlinespace 
syr2k & $N(N^2 + 2N + 1)/2$ & $\log(N) + 2$ & $2\log(N) + 2$\\ \addlinespace 
syrk & $N(N^2 + 2N + 1)/2$ & $\log(N) + 2$ & $2\log(N) + 2$\\ \addlinespace 
trisolv & $N(N + 3)/2$ & $N(\log(N - 1) + 3)$ & $N(\log(N - 1) + 4)$\\ \addlinespace 
trmm & $N^2(N + 3)/2$ & $\log(N - 1) + 3$ & $N^2(\log(N - 1) + 2)$\\

\bottomrule
\end{tabular}
\caption{Work-depth analysis of Polybench .}
\label{table:wd}
\end{table*}
\appendix

\section{Work and Depth}

In Table~\ref{table:wd} we list the Work and Depth of each Polybench application, using automatic parallelism extraction and extracted from manually-developed examples.

